\documentclass[12pt,oneside,fleqn]{article}
\usepackage{epsf}
\usepackage{amsfonts, bm}
\usepackage{amsmath}
\usepackage{amssymb}
\usepackage[latin2]{inputenc}
\usepackage[T1]{fontenc}
\usepackage{epsfig}
\usepackage{mathrsfs}
\usepackage{verbatim}

\newcommand{\pdn}{\mathcal{P}_1}
\newcommand{\pd}{\mathcal{P}}
\newcommand{\en}{\mathcal{E}_1}
\newcommand{\e}{\varepsilon}
\newcommand{\tr}{\text{tr}}
\newcommand{\g}{G_1\times G_2}

\newcommand{\I}{\text{id}}

\newcommand{\mh}{\mathcal{H}}

\newcommand{\bigtimes}{\mathop{\mathchoice%
{\smash{\vcenter{\hbox{\LARGE$\times$}}}\vphantom{\prod}}%
{\smash{\vcenter{\hbox{\Large$\times$}}}\vphantom{\prod}}%
{\times}%
{\times}
}\displaylimits}

\newtheorem{thm}{Theorem}[section]
\newtheorem{df}{Definition}[section]
\newtheorem{lem}{Lemma}[section]
\newtheorem{cor}{Corollary}[section]

\newlength{\dinwidth}
\newlength{\dinmargin}
\DeclareMathAlphabet{\scr}{U}{rsfs}{m}{n}

\begin{document}
\renewcommand\thefootnote{\fnsymbol{footnote}}
\begin{center}
{\large {\bf Entanglement of positive definite functions on compact groups}}\\
\vspace{0.5cm}
J. K. Korbicz$^{1,2}$\footnote{jaroslaw.korbicz@icfo.es}, J. Wehr$^{3,2}$,
and M. Lewenstein$^{2}$\\
\vspace{0.3cm}
$^1$ Dept. d'Estructura i Constituents de la Mat\`eria,
Universitat de Barcelona, 647 Diagonal, 08028 Barcelona, Spain \\

$^2$ ICREA and ICFO--Institut de Ci\`{e}ncies Fot\`{o}niques, Mediterranean Technology Park, 08860
Castelldefels (Barcelona), Spain\\

$^3$ Department of Mathematics, University of Arizona, 617 N. Santa Rita Ave.,
Tucson, AZ 85721-0089, USA

\end{center}
\setcounter{footnote}{0}
\renewcommand\thefootnote{\arabic{footnote}}

{\abstract We define and study entanglement of continuous positive
definite functions on products of compact groups. We formulate and
prove an infinite-dimensional analog of the Horodecki Theorem,
giving a necessary and sufficient criterion for separability of
such functions. The resulting characterisation is given in terms
of mappings of the space of continuous functions, preserving
positive definiteness. A relation between the developed
group-theoretical formalism and the conventional one, given in
terms of density matrices, is established through the
non-commutative Fourier analysis. It shows that the presented
method plays the role of a ``generating function'' formalism for
the theory of entanglement.}

\section{Introduction}

Entanglement is a property of states of composite quantum
mechanical systems. This concept lies at the very heart of quantum
mechanics, and it concerns all of the important aspects of quantum
theory: from philosophical aspects \cite{bell,malboro}, through
physical \cite{peres-book} and mathematical  \cite{mama, stroemer}
fundamentals\footnote{For a description of positive maps 
from a physical point of view, see e.g. Ref. \cite{maps}.}
, to applications in quantum information and metrology
\cite{Nielsen}. The importance of entangled states for the
understanding of quantum theory was recognized quite early, mainly
thanks to Einstein (e.g. in the famous EPR paper by Einstein,
Podolsky, and Rosen \cite{EPR}). Only with the advent of new
experimental techniques in recent years, it became clear that
entanglement may in fact also be used as a resource for
transmission and processing of (quantum) information, e.g. for
quantum cryptography or quantum computing (for a recent  review,
see Ref. \cite{mama} ; see also Ref. \cite{Nielsen}).

In the present work we develop a novel framework for
studying quantum entanglement, based on analysis
of continuous functions on compact groups.
With respect to the standard formalism of
entanglement theory, our approach plays a role
analogous to that of a ``generating function'' method---various
group-theoretical objects serve as ``generating functions''
for the corresponding families of operator-algebraic objects
(like density matrices, positive maps, etc), operating in different
dimensions.
This allows one to formulate and address the questions of entanglement
theory
in a unified, dimension-wise, way.

Before we proceed with the group-theoretical formalism, let us
first recall some basic facts and define the notion of
entanglement precisely. A quantum system is associated with  a
Hilbert space $\mh$, which we will assume to have a countable
basis. A state of the system is then represented by a positive,
trace-class operator $\varrho$ (a density matrix), satisfying
normalization condition $\tr\varrho=1$. If the system under
consideration is composite, i.e. it can be thought of being
composed of two subsystems $\mathscr A$ and $\mathscr B$, each of
which is treated as an independent individual, then, according to
the postulates of quantum theory, the Hilbert space of the system
is $\mh=\mh_{\mathscr A}\otimes \mh_{\mathscr B}$. The following
definition thus makes sense \cite{Werner}:

\begin{df}\label{sepdef}
A state $\varrho$ on $\mh_{\mathscr A}\otimes \mh_{\mathscr B}$ is called
separable if it can be approximated in the trace norm by convex combinations
of the form:
\begin{equation}
\sum_{m=1}^K p_m |x_m\rangle\langle x_m|\otimes |y_m\rangle\langle y_m|,
\ \ \text{where} \ \ x_m\in\mh_{\mathscr A},
y_m\in\mh_{\mathscr B},\
p_m\geqslant 0, \sum_{m=1}^K p_m=1.
\end{equation}
Otherwise $\varrho$ is called entangled.
\end{df}
This definition can be easily generalized to multipartite systems
with more than two parties involved.

In the light of the Definition \ref{sepdef} a
natural question arises, known as the {\it separability problem}:
Given a state
$\varrho$ decide if it is separable or not.

The problem turns out to be computationally very hard: although
efficient algorithms employing positive definite programming
methods exist  in lower dimensions \cite{Doherty}
(for a specific formulation of semi-definite approach for
$2\otimes N$ systems see Ref. \cite{woerdman}), it has been
proven that the problem belongs to the $NP$ complexity class as
dimensions of the Hilbert spaces involved grow \cite{gurvitz}. In
term of operational entanglement criteria up to date there are
only partial answers known, in both finite and infinite
dimensions. We briefly quote below  few basic results, referring
the reader to Ref. \cite{mama} for a complete overview. One
astonishingly powerful, given its simplicity, necessary criterion
for separability follows immediately from the definition of
separable states \cite{Peres,Horodeccy}:

\begin{thm}[Positivity of Partial Transpose (PPT)]\label{ppt}
If a state $\varrho$ on $\mathcal{H}_{\mathscr A}\otimes\mathcal{H}_{\mathscr B}$
is separable then the partially transposed operator
$\varrho^{T_{\mathscr B}}:=({\bf 1}_{\mathscr A}\otimes T)\varrho$ is positive,
where $T$ is a transposition map and ${\bf 1}_{\mathscr A}$ is the identity
operator on $\mh_{\mathscr A}$.
\end{thm}

In the lowest non-trivial dimensions
$\text{dim}\mh_{\mathscr A}=\text{dim}\mh_{\mathscr B}=2$ and
$\text{dim}\mh_{\mathscr A}=2$, $\text{dim}\mh_{\mathscr A}=3$
PPT criterion provides both necessary and sufficient
condition for separability
(see Ref. \cite{Horodeccy}, Theorem 3).
However, in higher dimensions
there exist states, called {\it PPT} or {\it bound entangled},
which satisfy the PPT criterion, but are nevertheless
entangled. The first examples of such states were
constructed in Ref. \cite{Choi_be} (although
in a different context of, so called, indecomposable maps)
and in Ref. \cite{Pawel}.

In infinite dimension, a complete solution to the separability
problem exists only for a special family of states---so called
Gaussian states \cite{Giedke}.

As mentioned above the separability problem is connected to other
open mathematical problems. In their fundamental work
\cite{Horodeccy} Horodecki {\it et al.} established an important
link between this problem and the problem of characterization of
positive maps on finite-dimensional matrix algebras
 (cf. Ref. \cite{Horodeccy},
Theorem 2; see also Refs. \cite{stroemer,Choi_be,Choi,Woronowicz} ):

\begin{thm}[M., P., and R. Horodecki]\label{hor}
Let $\mathcal{L}(\mathcal{H})$ denote the space of
linear operators on $\mathcal{H}$ and let
$\varrho\in\mathcal{L}(\mathcal{H}_{\mathscr A}\otimes\mathcal{H}_{\mathscr B})$
be a density matrix on a finite dimensional Hilbert
space $\mathcal{H}_{\mathscr A}\otimes\mathcal{H}_{\mathscr B}$.
Matrix $\varrho$ is separable if and only if for all linear maps
$\Phi\colon\mathcal{L}(\mathcal{H}_{\mathscr
B})\to\mathcal{L}(\mathcal{H}_{\mathscr A})$
preserving positive
operators (such maps are called positive), $({\bf 1}_{\mathscr
A}\otimes\Phi)\varrho\geqslant 0$ as an operator on $\mh_{\mathscr
A}\otimes\mh_{\mathscr A}$.
\end{thm}


The starting point for the present work is the non-commutative
Fourier analysis on a compact group, which we proposed to employ
for studying entanglement in Ref. \cite{terakurwamy} (cf. Ref.
\cite{FP} where the same method was used for a rigorous derivation
of the classical limit of quantum state space). Namely, one can
pass from operators $A\in\mathcal{L}(\mh_{\mathscr
A}\otimes\mh_{\mathscr B})$ to their non-commutative Fourier
transforms\footnote{We note that in Ref. \cite{Holevo} 
the term "noncommutative Fourier transform" is used in a slightly
different---though very closely related---sense.}
in two steps: i) identify the spaces
$\mh_\mathscr A$, $\mh_\mathscr B$ with representation spaces of
unitary, irreducible representations $\pi_\alpha$, $\tau_{\beta }$
of some compact groups $G_1$ and $G_2$ respectively (there are no
a priori restrictions on $G_1, G_2$ apart from possessing
representations in suitable dimensions); ii) pass from $A$ to a
function $\varphi_A\colon \g\to \mathbb C$, the non-commutative
Fourier transform of $A$, through:
\begin{equation}\label{glowny}
A\mapsto\varphi_A(g_1,g_2):=\text{tr}\big[A \pi_\alpha(g_1)\otimes\tau_{\beta }(g_2)\big].
\end{equation}
The above transform is called non-commutative, since apart from
the trivial case $\text{dim}\mh_{\mathscr
A}=\text{dim}\mh_{\mathscr B}=1$, groups $G_1$ and $G_2$ are
necessarily non-Abelian. In case $A=\varrho$ is a quantum state,
the corresponding function $\varphi_\varrho$ is called {\it
non-commutative characteristic function} of $\varrho$. The
transformation (\ref{glowny}) is invertible---one can recover $A$
from $\varphi_A$. Hence, one expects that for density matrices
their non-commutative characteristic functions should encode
entanglement in some way \cite{terakurwamy}. This is indeed the
case and in what follows we define and study the notion of
separability for suitably generalized non-commutative
characteristic functions (general continuous positive definite
functions on $\g$; cf. Definition \ref{do}). We then prove an
analog of the Horodecki Theorem \ref{hor} for such functions,
which constitutes the main result of the paper. Since the
framework we work in is countably infinite-dimensional (unless
both $G_1, G_2$ are finite) our result can be viewed as a
generalization of Horodecki Theorem to an infinite-dimensional
setting. The usual quantum-mechanical formalism, given by density
matrices, and the presented group-theoretical one are then shown
to be related through non-commutative harmonic analysis. In
particular, by employing non-commutative Fourier transform we
demonstrate how our approach turns out to be a ``generating
function'' method for the theory of entanglement.

Let us finally remark that the formalism of non-commutative
Fourier transform (\ref{glowny}) is closely related to that of
generalized coherent states \cite{Perelomov}. The difference is
that in the coherent state formalism one assigns to an operator
$A$ a function (called $P$-representation of $A$), which is
defined not on the whole group $G$, but on a homogeneous space
$G/H$, where $H$ is an isotropy subgroup of a fixed vector.
However, unlike non-commutative Fourier transform $\varphi_A$,
$P$-representation is generally non-unique (e.g. in $SU(2)$ case)
and does not encode positivity of a density matrix in a simple
manner. For some applications of generalized coherent states to
the study of entanglement see e.g. Refs. \cite{coh}.

\section{Preliminary notions}
In the main part of the work $G_1$, $G_2$ will be compact groups.
The principal object of our
study are continuous positive definite functions on the product
group $\g$. But first we
recall some basic definitions and facts, valid
for any locally compact $G$ (see e.g. Refs. \cite{Folland,Dixmier,HewittII}
for a complete exposition).

\begin{df}\label{do}
A continuous complex function $\varphi$ on a group $G$ with the
Haar measure $dg$ is called positive definite if it is bounded and
satisfies:
\begin{equation}
\iint d g dh \overline{f(g)}\varphi(g^{-1}h)f(h)\geqslant 0\label{PD}
\end{equation}
(bar denotes complex conjugation) for any continuous function $f$ with compact support.
\end{df}

We will denote by $\mathcal{P}(G)$ the set of positive definite functions on $G$
and by $\pdn(G)$ its subset consisting of the functions which satisfy the normalization $\varphi(e)=1$,
where $e$ is the neutral element of $G$.
$\pd(G)$ is a closed convex cone in $C(G)$ --- the space of continuous complex-valued functions on
$G$ equipped with the topology of uniform convergence on compact sets, called compact convergence in the sequel.

The structure of $\mathcal{P}(G)$ is described by the following
deep, fundamental result of representation theory, often referred
to as the GNS construction (see e.g. Ref. \cite{Folland}, Theorem
3.20; Ref. \cite{Dixmier}, Theorem 13.4.5):

\begin{thm}[Gel'fand, Naimark, Segal]\label{GNS}
With every  $\varphi\in\pd(G)$ we can associate a Hilbert space
$\mathcal{H}_\varphi$, a unitary representation  $\pi_\varphi$ of
$G$ in $\mathcal{H}_\varphi$ and a vector $v_\varphi$, cyclic for
$\pi_\varphi$, such that:

\begin{equation}\label{GNSeq}
\varphi(g)=\langle v_\varphi |\pi_\varphi (g) v_\varphi\rangle.
\end{equation}
The representation $\pi_\varphi$ is unique up to a unitary equivalence.
\end{thm}

The above result provides a tool for a systematic study of
$\mathcal{P}(G)$ in terms of representations of $G$ (and
conversely). In the sequel we will need some basic properties of
positive-definite functions.  While they all follow from the
definition by standard, elementary arguments, we find the proofs
based on the GNS representation particularly transparent.

A function $\varphi$ will be called {\it pure} if $\pi_\varphi$ is
irreducible. Pure normalized functions are the extreme points of
$\pdn(G)$ (cf. Ref. \cite{Folland}, Theorem 3.25); we denote their
set by $\en(G)$. Every $\varphi\in\pdn(G)$ is a limit, in the
topology of compact convergence, of convex combinations of extreme
points of $\pdn(G)$ (cf. Ref. \cite{Dixmier}, Theorem 13.6.4):
\begin{equation}
g\mapsto \sum_{m=1}^N p_m\e_m(g), \ \ \text{where}\ \ \e_m\in\en(G),\
p_m\geqslant 0,\ \sum_{m=1}^N p_m=1.
\end{equation}
There is also an integral representation (provided $G$ is
separable as a topological space), sometimes called generalized
Bochner Theorem. Namely, for any $\varphi\in\pdn(G)$ there exists
a probability measure $\mu_\varphi$ concentrated on $\en(G)$ such
that (cf. Ref. \cite{Dixmier}, Proposition 13.6.8):
\begin{equation}\label{intrep}
\varphi(g)=\int\limits_{\en(G)}d \mu_\varphi(\e)\,\e(g) \ \ \ \text{for any}\  g\in G.
\end{equation}

From this point on, we assume that $G_1, G_2$ are {\it compact}
and consider positive definite functions on $\g$. Let us introduce
the algebraic tensor product $C(G_1)\otimes C(G_2)$ as the space
of finite (complex) linear combinations of product functions
$f\otimes\xi\colon(g_1,g_2)\mapsto f(g_1)\xi(g_2)$. Then
$C(G_1)\otimes C(G_2)$ is uniformly dense in $C(\g)$.  This
standard fact follows, for example, from the Stone-Weierstrass
Theorem (see e.g. Ref. \cite{Folland_anal}).  Every product
$\phi\otimes\psi$, where $\phi\in\pdn(G_1),\psi\in\pdn(G_2)$, is
positive definite on $\g$, since, by the GNS Theorem \ref{GNS},
$\phi(g_1)\psi(g_2) = \langle v_\phi\otimes v_\psi
|\pi_\phi(g_1)\otimes\pi_\psi(g_2) v_\phi\otimes v_\psi\rangle$
which is of the form (\ref{GNSeq}) on $\g$. It follows that convex
combinations of such products are positive definite and hence so
are uniform limits of such convex combinations. The resulting
class of positive definite functions, introduced formally in the
next definition, is our fundamental object of study (compare
Definition \ref{sepdef}).

\begin{df}\label{Sep}
We define $Sep_0$ as the set of all functions $\varphi\in\pdn(\g)$
which can be represented as finite convex combinations
\begin{equation}\label{sep}
\varphi(g_1,g_2)=\sum_{m=1}^K p_m \e_m(g_1)\eta_m(g_2), \ \ \text{where}\
\e_m\in\en(G_1),\eta_m\in\en(G_2).
\end{equation}
A function $\varphi\in\pdn(\g)$ is called separable if
it is a uniform limit of elements of $Sep_0$.
The set of separable functions is denoted by $Sep$.
Functions which are not separable are called entangled.
\end{df}

The definitions of separable and entangled functions
generalize without any change to arbitrary (i.e. not necessarily normalized) positive definite
functions. This includes our main result, Theorem \ref{main}, together
with its proof (since for a nonzero positive definite function
$\varphi(e_1,e_2)=||\varphi||_\infty> 0$, we can replace $\varphi$ by
$\varphi/\varphi(e_1,e_2)$ and reduce the proof to the normalized case).
The normalization is, however,
natural from the physical point of view.

Geometrically $\en(G_1)\times \en(G_2)$ is embedded into $\en(\g)$ through the map
$(\e,\eta)\mapsto \e\otimes\eta$. Then $Sep$ is a closed convex hull of
$\en(G_1)\times \en(G_2)$.

We note that every $\varphi\in Sep$ admits an integral representation:
\begin{equation}
\varphi(g_1,g_2)=\int\limits_{\en(G_1)\times \en(G_2)}d \mu_\varphi(\e,\eta)\,
\e(g_1)\eta(g_2) \ \ \ \text{for any}\  (g_1,g_2)\in \g,
\end{equation}
but we will not use this fact.

\section{Necessary and sufficient criterion for separability
of positive definite functions}\label{GrHor} The main problem we
would like to address is that of finding an {\it intrinsic}
characterization of separable functions $\varphi\in Sep$. This is
known as {\it the generalized separability problem}
\cite{terakurwamy}. By Eq. (\ref{GNSeq}) for every positive
definite function $\phi$, $\phi(g^{-1})=\overline{\phi(g)}$ and
this function is again positive definite. It now follows
immediately from Definition \ref{Sep} and from uniform closedness
of $\pd(\g)$ in $C(\g)$ that:

\begin{thm}\label{PPT}
If $\varphi\in Sep$ then
the function $(g_1,g_2)\mapsto\varphi(g_1,g_2^{-1})$
is positive definite.
\end{thm}

The above simple criterion is only a necessary condition---there
are functions satisfying it which are nevertheless entangled. This
can be seen by noting that Theorem \ref{PPT} is a
group-theoretical analog of the PPT criterion, given  by Theorem
\ref{ppt}, (we will show it in Section \ref{examples}; see also
Ref. \cite{terakurwamy}, Theorem 2) and, as we mentioned in the
Introduction, there exist PPT entangled (or, equivalently, bound
entangled) quantum states \cite{Pawel}. A natural question arises
whether one obtains a complete characterization of separable
functions when in place of the inverse $g \mapsto g^{-1}$ one
considers all possible linear maps of functions, preserving
positive definiteness. The affirmative answer is the main result
of our work:

\begin{thm}\label{main}
A function $\varphi\in\pdn(\g)$ is separable if and only
if for every bounded linear map
$\Lambda\colon C(G_2)\to C(G_1)$, such that $\Lambda\pd(G_2)\subset\pd(G_1)$, function
$(\I\otimes\Lambda)\varphi$ is positive definite on $G_1\times G_1$.
\end{thm}

Tensor product $\I\otimes\Lambda\colon C(G_1\times G_2)\to
C(G_1\times G_1)$ is defined in the natural way: we first define
it on the algebraic product $C(G_1)\otimes C(G_2)$ and then extend
by continuity to all of $C(G_1\times G_2)$.

The above theorem is a group-theoretical analog of the Horodecki Theorem
\ref{hor}.  In fact, we derive a version of the Horodecki result
as a corollary in Section \ref{ConHor} (cf. Theorem \ref{grouphor}).
We will adopt the standard terminology of entanglement theory and
say that an entangled function $\varphi$ is {\it detected by a map
$\Lambda$} if the function $(\I\otimes\Lambda)\varphi$ is not positive definite.

As mentioned earlier, the above theorem, as well as the following
proof, hold for arbitrary positive definite functions, but in
order to use more natural concepts from the physical point of view
we state and prove it for normalized ones. The proof in one
direction is immediate and follows directly from the Definition
\ref{Sep}: $(\I\otimes\Lambda)\sum_{m=1}^K p_m \e_m\otimes\eta_m
=\sum_{m=1}^K p_m\,\e_m\otimes\Lambda\eta_m\in\pd(G_1\times G_1)$
and since $\pd(G_1\times G_1)$ is uniformly closed in $C(G_1\times
G_1)$ this holds on all of $Sep$.

For the proof in the other direction, let $C(\g)'$ denote the
space of continuous linear functionals on $C(\g)$ (the space dual
to $C(\g)$).  Since $Sep$ is a closed convex set, it follows from
the Hahn-Banach Theorem (see e.g. Ref. \cite{ReedI}, Theorem V.4)
that for every $\varphi\notin Sep$ there exists a functional $l\in
C(\g)'$ and a real number $\gamma$, such that:
\begin{equation}\label{hb1}
\text{Re} l(\varphi)<\gamma\leqslant\text{Re}l(\sigma) \ \ \ \text{for any}\
\sigma\in Sep,
\end{equation}
where $\text{Re}l$ denotes the real part of the functional $l$:
$\text{Re}l(\varphi):= \text{Re}[l(\varphi)]$. From the Riesz
Representation Theorem (see e.g Ref. \cite{ReedI}, Theorem IV.17)
we know that each linear functional $l$ on $C(\g)$ can be uniquely
represented by a complex measure $\mu_l$ with finite total
variation $|\mu_l|$ on $\g$. Denoting the space of such measures
by $M(\g)$ we have: $C(\g)'=M(\g)$. We will interchangeably treat
elements of $C(\g)'$ as either linear functionals or as the
corresponding measures. To work with the normalized functions
$\varphi$ which we are interested in, it is convenient to
introduce a modification of the functional $l$ in (\ref{hb1}):
$L:=l-\gamma\delta_{(e_1,e_2)}$, where $\delta_{(e_1,e_2)}$ is the
Dirac delta (point mass), concentrated at the neutral element
$(e_1,e_2)$ of $\g$. Hence, for every entangled
$\varphi\in\pdn(\g)$ there exists $L\in C(\g)'$ such that:
\begin{equation}\label{hb2}
\text{Re}L(\varphi)<0\leqslant\text{Re}L(\sigma) \ \ \ \text{for any}\
\sigma\in Sep.
\end{equation}
As an easy consequence of the above condition we obtain the following lemma,
crucial for the rest of the proof:

\begin{lem}\label{HB}
A function $\varphi$ is separable if and only if for every functional $L\in C(\g)'$, satisfying
$\text{Re}L(\psi_1\otimes\psi_2)\geqslant 0$ for every $\psi_1\in\pdn(G_1),\psi_2\in\pdn(G_2)$,
we have $\text{Re}L(\varphi)\geqslant 0$.
\end{lem}

Indeed, for every $Sep_0\ni\varphi=\sum_{m=1}^K p_m
\e_m\otimes\eta_m$, $L(\varphi)=\sum_{m=1}^K p_m
L(\e_m\otimes\eta_m)\geqslant 0$ and by continuity of $L$ this
extends to all of $Sep$. Conversely, assume that for every $L$
satisfying the condition in the statement of the lemma,
$\text{Re}L(\varphi)\geqslant 0$ but $\varphi\notin Sep$. Then
from the Hahn-Banach Theorem (see \ref{hb2}) we know that there
exists a functional $L_0$ such that $\text{Re}L_0(\sigma)\geqslant
0$ for every separable $\sigma$---in particular for every function
of the form $\psi_1\otimes\psi_2$---and $\text{Re}L_0(\varphi)<
0$, which contradicts our assumption.$\Box$
\\

In order to pass from linear functionals on $C(\g)$ to linear maps
from $C(G_2)$ to  $C(G_1)$, we first employ an algebraic
isomorphism between functionals from $C(\g)'$ and bounded linear maps
$\widetilde\Lambda\colon C(G_2)\to C(G_1)'$. For each $L\in
C(\g)'$ we define the corresponding map $\widetilde\Lambda_L$ by:
\begin{equation}\label{izojam}
\widetilde\Lambda_L\xi(f):=L(f\otimes \xi)
\end{equation}
$\widetilde\Lambda_L$ is bounded:
\begin{equation}\label{bound}
||\widetilde\Lambda_L||=\sup_{||\xi||_\infty=1}||\widetilde\Lambda_L\xi||_\infty'
=\sup_{||\xi||_\infty=1}\,\sup_{||f||_\infty=1}|\widetilde\Lambda_L\xi(f)|
=\sup_{||\xi||_\infty=1}\,\sup_{||f||_\infty=1}|L(f\otimes\xi)|,
\end{equation}
where $||\cdot ||_\infty'$ is the norm on $C(G_1)'$ induced by the
supremum norm $||\cdot ||_\infty$. Conversely, for any bounded map
$\widetilde\Lambda\colon C(G_2)\to C(G_1)'$, Eq. (\ref{izojam})
defines a functional $L_{\widetilde\Lambda}$ on $C(G_1)\otimes
C(G_2)$, which is bounded by Eq. (\ref{bound}), and uniquely
defined, since if $L_{\widetilde\Lambda}(f\otimes \xi)=0$ for all
$f\in C(G_1),\xi\in C(G_2)$, then by Eq. (\ref{izojam})
$\widetilde\Lambda\equiv 0$ (note that $L_{\widetilde\Lambda}$
depends linearly on $\widetilde\Lambda$) . As  $C(G_1)\otimes
C(G_2)$ is uniformly dense in $C(\g)$, $L_{\widetilde\Lambda}$ can
be uniquely extended to a continuous functional on all of $C(\g)$.
This establishes the claimed isomorphism $L \leftrightarrow
\widetilde\Lambda_L$.


Next, we establish a positivity criterion $\widetilde\Lambda_L$, analogous to the one
given by Jamio\l kowski in Ref. \cite{Jamiolkowski} for operators on finite
dimensional Hilbert spaces:

\begin{lem}\label{jam}
A functional $L\in C(\g)'$ satisfies $\text{Re}L(\psi_1\otimes\psi_2)\geqslant 0$
for all $\psi_1\in\pdn(G_1),\psi_2\in\pdn(G_2)$
if and only if $\text{Re}\widetilde\Lambda_L$ maps positive definite functions
from $\pd(G_2)$ to positive
definite measures from $M(G_1)$.
\end{lem}

Positive definite measure on $G$ is a measure $\mu$ satisfying a
generalization of the condition (\ref{PD})\footnote{On an arbitrary
locally compact $G$ positive definite measures are defined by requiring that
condition (\ref{pdm}) hold for all continuous functions with compact support
(cf. Ref. \cite{Dixmier}, Definition 13.7.1).}:
\begin{equation}\label{pdm}
\int d\mu\,(f^*\ast f)\geqslant 0 \ \ \ \text{for any}\  f\in C(G),
\end{equation}
where $f^*(g):=\overline{f(g^{-1})}$ is the involution
and $(f\ast \xi)(h):=\int d g f(g)\xi(g^{-1}h)$ is the
convolution. The action of the map
$\text{Re}\widetilde\Lambda_L$ on $f$ is defined as the
real part of the functional $\widetilde\Lambda_L f$.
The condition $\text{Re}L(\psi_1\otimes\psi_2)\geqslant 0$
for all normalized $\psi_1\in\pdn(G_1),\psi_2\in\pdn(G_2)$ is equivalent
to $\text{Re}L(\psi_1\otimes\psi_2)\geqslant 0$
for all $\psi_1\in\pd(G_1),\psi_2\in\pd(G_2)$, since we can
replace $\psi_1,\psi_2\ne 0$ by $\psi_1/\psi_1(e)$
and $\psi_2/\psi_2(e)$. From the definition
(\ref{izojam}) it follows that:
\begin{equation}\label{izojam2}
\text{Re}L(\psi_1\otimes\psi_2)=\text{Re}[\widetilde\Lambda_L\psi_2(\psi_1)]
=\text{Re}\widetilde\Lambda_L\psi_2(\psi_1).
\end{equation}
A theorem by Godement (cf. Ref. \cite{Godement}, Theorem 17; Ref.
\cite{Dixmier}, Theorem 13.8.6) states that every positive
definite function can be uniformly approximated by functions of
the form $f^*\ast f$ where $f$ is continuous with compact support.
Hence, since $G_1, G_2$ are compact,
$\text{Re}L(\psi_1\otimes\psi_2)\geqslant 0$ for every
$\psi_1\in\pd(G_1),\psi_2\in\pd(G_2)$ if and only if
$\text{Re}L\big[(f^*\ast
f)\otimes\psi_2\big]=\text{Re}\widetilde\Lambda_L\psi_2(f^*\ast f)
\geqslant 0$ for every $f\in C(G_1)$ and $\psi_2\in\pd(G_2)$. But
by Eqs. (\ref{pdm}) and (\ref{izojam2}) this is equivalent to the
measure $\text{Re}\widetilde\Lambda_L\psi_2\in M(G_1)$ being
positive definite for every $\psi_2\in\pd(G_2)$.$\Box$
\\

As the next step we will regularize maps $\widetilde\Lambda$.
Note that for any $\mu\in M(G_1)$ and
$f\in C(G_1)$ the convolution:
\begin{equation}\label{splot1}
\mu\ast f (h) = \int d\mu(g) f(g^{-1}h)
\end{equation}
is a continuous function on $G_1$. Let $\{\psi_U\}$, where
$U\subset G_1$ runs through a neighbourhood base of the neutral
element $e_1\in G_1$, be an approximate identity in $C(G_1)$. That
is, for every $U$ we have:
\begin{eqnarray}
& &o)\; \psi_U\in C(G_1),\quad
i)\; \text{supp}\psi_U\ \text{is a compact subset of}\ U,\label{prop1}\\
& &ii)\; \psi_U\geqslant 0,\quad iii)\; \psi_U(g^{-1})=\psi_U(g),\quad
iv)\; \int \psi_U=1.\label{prop2}
\end{eqnarray}
Using definition (\ref{splot1}), let us define for every $f\in
C(G_1)$ functions
\begin{equation}\label{LU}
\Lambda_U f := \widetilde\Lambda f\ast \psi_U.
\end{equation}
Then $\Lambda_U f$ converges in weak-$\ast$ topology to
$\widetilde\Lambda f$ as $U \to \{e_1\}$. To see this, let us
calculate $\int dg \Lambda_U f (g)\xi(g)$ for an arbitrary $\xi\in
C(G_1)$:
\begin{eqnarray}
\int d g \,(\widetilde\Lambda f \ast \psi_U)(g) \xi(g) &=&
\int d g \int d (\widetilde\Lambda f)(h) \psi_U(h^{-1}g)\xi(g)
= \int d (\widetilde\Lambda f)(h)  \int d g \psi_U(g^{-1}h)\xi(g)\nonumber\\
&=& \widetilde\Lambda f (\xi\ast\psi_U),
\end{eqnarray}
where in the second step we used the symmetry of $\psi_U$: $\psi_U(g^{-1})=\psi_U(g)$.
But from the properties (\ref{prop1}), (\ref{prop2}) of $\{\psi_U\}$ it follows that
$\xi\ast\psi_U \to \xi$ uniformly as $U \to \{e_1\}$
(cf. Ref. \cite{Folland}, Theorem 2.42), which proves
the desired weak-$\ast$ convergence.
Thus any bounded map $\widetilde\Lambda\colon C(G_2)\to M(G_1)$
can be weakly-$\ast$ approximated
by bounded maps $\Lambda_U\colon C(G_2)\to C(G_1)$ (boundedness of $\Lambda_U$
for every $U\subset G_1$ follows immediately from the
definitions (\ref{splot1}) and (\ref{LU})).

In order to preserve the positivity property of the
$\widetilde\Lambda$'s, introduced in Lemma \ref{jam}, we choose
regularizing functions $\psi_U$ in a special way. Namely, for
every neighbourhood $U$ of $e_1\in G_1$ we can find such open
$V\ni e_1$  that: (cf. Ref. \cite{Folland}, Lemma 5.24):
\begin{equation}\label{V}
o)\; V\subset U,\quad i)\; V^{-1}=V,\quad ii)\; gVg^{-1}=V\ \text{for every}\ g\in G_1.
\end{equation}
Let us define the functions:
\begin{eqnarray}
\kappa_V &:=& \frac{1}{|V|}\chi_V\label{al}\\
\psi_U &:=& \kappa_V\ast\kappa_V,\label{reg}
\end{eqnarray}
where $\chi_V$ is the indicator function of $V$ and $|V|=\int dg
\chi_V(g)$ is the Haar measure of $V$. Then one easily shows that
$\{\psi_U\}$ form an approximate identity in $C(G_1)$. Moreover,
$\kappa_V$, and hence $\psi_U$, are central functions, i.e. for
every $g$ and $h$, $\kappa_V(gh)=\kappa_V(hg)$, which follows from
the property $(ii)$ of the sets $V$. Using functions (\ref{reg}) to
regularize an arbitrary map $\widetilde\Lambda$, we find that:
\begin{equation}\label{pdreg}
\Lambda_U f = \widetilde\Lambda f\ast \psi_U
= \kappa_V\ast\widetilde\Lambda f\ast\kappa_V=\kappa_V^*\ast\widetilde\Lambda f\ast\kappa_V,
\end{equation}
where in the second step we used the fact that $\kappa_V$ are central and
hence $\mu\ast\kappa_V=\kappa_V\ast\mu$ for any $\mu$. In the last step we used the symmetry
condition $(i)$ and $\overline{\chi_V}=\chi_V$. Now, from the
special form of the regularization (\ref{pdreg}) we obtain:

\begin{lem}\label{mainreg}
If $\widetilde\Lambda\colon C(G_2)\to M(G_1)$ maps positive
definite functions into positive definite measures, then
for every neighbourhood $U$ of $e_1\in G_1$
the regularized maps $\Lambda_U$, defined by Eqs. (\ref{al}-\ref{pdreg}),
map $\pd(G_2)$ into $\pd(G_1)$.
\end{lem}

To prove Lemma \ref{mainreg},
note that for an arbitrary $\phi\in\pd(G_2)$ and $f\in C(G_1)$ one has:
\begin{eqnarray}
& & \iint dg dh \,\overline{f(g)}\Big(\kappa_V^*\ast\widetilde\Lambda \phi\ast\kappa_V\Big)(g^{-1}h)
f(h) \nonumber\\
& & =\iiint dg dh da \,\overline{f(g)}\;\overline{\kappa_V(a^{-1})}
\int d (\widetilde\Lambda \phi)(b) \,\kappa_V(b^{-1}a^{-1}g^{-1}h)f(h)\nonumber\\
& & = \int d(\widetilde\Lambda \phi)(b)\int da \int dg \,\overline{f(g)}
\;\overline{\kappa_V(ag)}\int dh \,f(h)\kappa_V(b^{-1}ah)\nonumber\\\
& & = \int d(\widetilde\Lambda \phi)(b)\int da \,\big(f\ast\check\kappa_V\big)^*(a)
\big(f\ast\check\kappa_V\big)(a^{-1}b)=\widetilde\Lambda \phi\Big[(f\ast\kappa_V\big)^*
\ast(f\ast\kappa_V)\Big],\label{pdreg2}
\end{eqnarray}
where $\check\kappa_V(g):=\kappa_V(g^{-1})=\kappa_V(g)$
by property $(i)$ in Eq. (\ref{V}).
Hence, if $\widetilde\Lambda\phi$ is a positive definite
measure from $M(G_1)$ then for every $U$,
$\Lambda_U\phi$ is a positive definite function on $G_1$
(note that $f\ast\kappa_V$ is continuous since $f$ is).$\Box$
\\

Let us introduce some terminology, analogous to that used in the theory of
linear mappings of operators on
finite-dimensional Hilbert spaces
(see e.g. Refs. \cite{Kraus, Woronowicz, Choi}):

\begin{df}\label{posdef}
Let $\Lambda\colon C(G_2)\to C(G_1)$ be a bounded linear map.
Then $\Lambda$ is called:
\begin{itemize}
\item positive definite (PD)
if it preserves positive definite functions, i.e. if
$\Lambda\pd(G_2)\subset \pd(G_1)$;
\item $H$-positive definite ($H$-PD), where $H$ is a compact group,
if $\I\otimes\Lambda\colon
C(H\times G_2)\to C(H \times G_1)$
is positive definite, i.e. if
$\big(\I\otimes\Lambda\big)\pd(H\times G_2)\subset \pd(H\times G_1)$;
\item completely positive definite (CPD) if it is $H$-positive definite for any compact $H$.
\end{itemize}
\end{df}


Thus, rephrased in the terms introduced above,
Lemma \ref{mainreg} states that every bounded linear map
$\widetilde\Lambda\colon C(G_2)\to M(G_1)$, mapping $\pd(G_2)$
into positive definite measures, can be weakly-$\ast$
approximated by positive definite maps from $C(G_2)$ to $C(G_1)$.

After we have established almost all the necessary facts, we return
to the main Lemma \ref{HB}. First, rewrite Eq. (\ref{izojam})
using the regularization (\ref{mainreg}) in order to be able to
write down explicitly the right hand side of Eq. (\ref{izojam})
for arbitrary functions, not only product ones. For product functions
we have:
\begin{equation}
\text{Re}L(f\otimes\xi)=\text{Re}\widetilde\Lambda_L\xi(f).
\end{equation}
We apply the regularization (\ref{pdreg}) to $\text{Re}\widetilde\Lambda_L$,
denoting the regularized operators by $R_U$, $R_U\colon C(G_2)\to C(G_1)$, rather than by
$\big(\text{Re}\Lambda_L\big)_U$, to obtain:
\begin{eqnarray}
\text{Re}L(f\otimes\xi)=\lim_{U\to \{e_1\}}R_U\xi(f)
&=&\lim_{U\to \{e_1\}}\int\limits_{G_1} dg_1 \big(R_U\xi\big)(g_1) f(g_1)\nonumber\\
&=&\lim_{U\to \{e_1\}}\int\limits_{G_1} dg_1
\Big[\big(\I\otimes R_U\big)f\otimes\xi\Big](g_1,g_1)\label{wkoncukurwa}
\end{eqnarray}
Formula (\ref{wkoncukurwa}) immediately extends to all of $C(\g)$,
so in particular for any $\varphi\in\pdn(\g)$ we have:
\begin{equation}\label{finalstep}
\text{Re}L(\varphi)=
\lim_{U\to \{e_1\}}\int\limits_{G_1} dg_1 \Big[\big(\I\otimes R_U\big)\varphi\Big](g_1,g_1).
\end{equation}

Summarizing, by application of Lemmas \ref{jam} and \ref{mainreg}
we obtain that for an arbitrary
functional $L\in C(\g)'$, such that
$\text{Re}L(\psi_1\otimes \psi_2)\geqslant 0$ for every $\psi_{1},\psi_{2}\in\pdn(G)$,
$\text{Re}L(\varphi)$ is given by Eq. (\ref{finalstep}), where for every
neighbourhood $U\subset G_1$ the maps $R_U\colon C(G_2)\to C(G_1)$
are bounded and positive definite. We need two more
simple facts.

First, we note that if $\varphi$ is a positive definite function
on the product of two copies of $G_1$, i.e.
$\varphi\in\pd(G_1\times G_1)$, then its restriction to the
diagonal $\varphi\big|_\Delta(g):=\varphi(g,g)$ is a positive
definite function on $G_1$. A particularly direct proof of this
fact follows from the GNS construction (cf. Theorem \ref{GNS}):
\begin{eqnarray}
\iint dg dh\, \overline{f(g)}\varphi\big|_\Delta(g^{-1}h)f(h)&=&
\iint dg dh\, \overline{f(g)}\langle v_{\varphi}|
\pi_{\varphi}(g^{-1},g^{-1})
\pi_{\varphi}(h,h)\, v_{\varphi}\rangle f(h)\nonumber\\
&=& \bigg\langle\int dg f(g)\pi_{\varphi}(g,g) v_{\varphi}\bigg|
\int dh f(h)\pi_{\varphi}(h,h) v_{\varphi}\bigg\rangle\geqslant 0.
\end{eqnarray}

Second, we note that since $G_1$ is compact, $\int
dg\,\phi(g)\geqslant 0$ for any $\phi\in\pd(G_1)$ (cf. Ref.
\cite{HewittII}, Theorem 34.8). Indeed, if $G_1$ is compact then
the constant function $1$ is a function with compact support and
we can use it in the condition (\ref{PD}), which then implies that
$\iint dg dh\,\phi(g^{-1}h)=\int dh\,\phi(h)\geqslant 0$, where we
changed variables $h\mapsto gh$ and used the normalization of
$dg$.

To finish the proof of the main Theorem \ref{main}, let us assume
that for every bounded and positive definite map $\Lambda\colon
C(G_2)\to C(G_1)$, the function $(\I\otimes\Lambda)\varphi$ is
positive definite. It then follows from the above discussion and
from Eq. (\ref{finalstep}) that $\text{Re}L(\varphi)\geqslant 0$
for every $L$, such that $\text{Re}L(\psi_1\otimes\psi_2)\geqslant
0$ for every $\psi_1\in\pdn(G_1),\psi_2\in\pdn(G_2)$. But then
Lemma \ref{HB} implies that $\varphi\in Sep$.$\Box$

\section{Fourier transforms and ``generating function'' formalism}\label{ConHor}
In this section we establish a connection between the formalism of
positive definite functions and standard notions of entanglement
theory, thus ascribing a ``physical meaning'' to the former.
Namely, as advertised in the Introduction, we show that various
group-theoretical objects studied in the previous sections turn
out to be ``generating functions'' for the corresponding
operator-algebraic objects. For example a positive definite
function generates a family of (subnormalized) density matrices, a
positive definite map (cf. Definition \ref{posdef}) generates a
family of positive maps, etc. We also derive here a weaker version
of the Horodecki Theorem (cf. Theorem \ref{hor}) from Theorem
\ref{main} and prove a number of other useful results.

Our main tool will be non-commutative Fourier analysis of
continuous functions on compact groups.
Below we recall some basic notions and methods,
which we will need
(see e.g. Refs. \cite{Folland,Dixmier} for more).
The goal which we have in mind is to construct uniformly convergent
Fourier series for continuous functions.

By (a part of) the fundamental theorem of the theory---the
Peter-Weyl Theorem (see e.g. Ref. \cite{Folland}, Theorem 5.12),
any continuous function on a compact group can be uniformly
approximated by linear combinations of matrix elements of
irreducible representations, taken in some fixed orthonormal bases
of the corresponding representation spaces. Here, we are primarily
interested in product groups $\g$, so we first recall their
representation structure. Let us denote by $\widehat G_1$
$(\widehat G_2)$ the set of equivalence classes of irreducible,
strongly continuous, unitary representations ({\it irreps}) of
$G_1$ $(G_2)$. Since $G_1, G_2$ are compact, $\widehat G_1$ and
$\widehat G_2$ are discrete and (the classes of equivalent)
irreducible representations can be labelled by discrete indices.
We will denote irreps of $G_1$ and $G_2$ by $\pi_\alpha$ and
$\tau_\beta$ respectively and the spaces where they act by
$\mh_\alpha$ and $\widetilde\mh_\beta$. It can be then shown that
for a large family of groups, including compact ones, every irrep
of the product $\g$ can be chosen in the form
$\pi_\alpha\otimes\tau_{\beta }$, where:
\begin{equation}
\pi_\alpha\otimes\tau_{\beta }(g_1,g_2):=\pi_\alpha(g_1)\otimes \tau_\beta(g_2)
\end{equation}
acts in the space $\mh_\alpha\otimes \widetilde\mh_\beta$
(cf. Ref. \cite{Folland}, Theorem 7.25; Ref. \cite{Dixmier}, Proposition 13.1.8).
In other words,  $\widehat{\g}$ can be identified with
$\widehat G_1\times\widehat G_2$.

Next, we need matrix elements of the irreps
$\pi_\alpha\otimes\tau_{\beta }$. It is natural here to take them
with respect to product bases of the representation spaces
$\mh_\alpha\otimes\widetilde\mh_\beta$. Thus,  for each pair of
the representation indices $\alpha$ and $\beta$ we fix an
orthonormal base $\{e_i\}_{i=1,\dots,\text{dim}\mh_\alpha}$ of
$\mh_\alpha$ and an orthonormal base $\{\tilde
e_k\}_{k=1,\dots,\text{dim}\widetilde\mh_\beta}$ of
$\widetilde\mh_\beta$ (we do not indicate explicitly the
dependence of $\{e_i\}$, $\{\tilde e_k\}$ on the representation
indices $\alpha$, $\beta $ in order not to complicate the
notation). The corresponding matrix elements of
$\pi_\alpha\otimes\tau_\beta$ are then simply given by products of
the matrix elements of $\pi_\alpha$ and $\tau_\beta$---that is,
they are given by the functions $\pi^\alpha_{ij}\otimes\tau^{\beta
}_{kl}$, where:
\begin{equation}
\pi^\alpha_{ij}(g_1):=\langle e_i|\pi_\alpha(g_1)e_j\rangle,\quad
\tau^{\beta }_{kl}(g_2):=\langle\tilde e_k|\tau_{\beta }(g_2)\tilde e_l\rangle.
\end{equation}
Now, for a given $f\in C(\g)$
we can formally write the Fourier series:
\begin{equation}\label{fourier}
f=\sum_{\alpha,\beta }\sum_{i,\dots,l}f^{ijkl}_{\alpha\beta }\;
\pi^\alpha_{ij}\otimes\tau^{\beta }_{kl}, \quad
f^{ijkl}_{\alpha\beta }:=n_\alpha m_\beta\iint\limits_{G_1\ G_2} dg_1dg_2
\overline{\pi^\alpha_{ij}(g_1)}\:\overline{\tau^{\beta }_{kl}(g_2)}f(g_1,g_2),
\end{equation}
where $n_{\alpha}:=\text{dim}\mh_{\alpha},
m_\beta:=\text{dim}\mh_{\beta}$. However, for a generic function
$f\in C(\g)$ the Fourier series (\ref{fourier}) converges only in
the $L^2$ norm (since $\g$ is compact $C(\g)\subset L^2(\g)$) and
not uniformly.

The standard way around this difficulty is the following:
i) regularize $f$ so that the Fourier series
of the regularized function converges uniformly;
ii) perform the desired manipulations with the series;
iii) at the end uniformly remove
the regularization.
For the regularization the same technique
as in Section \ref{GrHor} (cf. Eqs. (\ref{prop1}-\ref{LU})
and Eqs. (\ref{V}-\ref{pdreg})) is used. Thus, for a given $f$ we
consider a function $f_{\mathcal U}:=f\ast\psi_{\mathcal U}$,
where
the regularizing functions $\psi_{\mathcal U}\in C(\g)$ are defined in an analogous way
as in Eqs. (\ref{V}-\ref{reg}), but this time on the product $\g$.
The sets $\mathcal U\subset\g$ now
run through a neighborhood base of the neutral element
$\{e_1,e_2\}\in\g$.
Since, by construction, the
$\psi_{\mathcal U}$'s are central on $\g$ (cf. property $(ii)$
in Eq. (\ref{V})), a simple calculation shows that
the Fourier series of $f\ast\psi_{\mathcal U}$ takes the following form:
\begin{equation}\label{fourierU}
f\ast\psi_{\mathcal U}=\sum_{\alpha,\beta }\sum_{i,\dots,l}
c_{\mathcal U}^{\alpha\beta }f^{ijkl}_{\alpha\beta }\;
\pi^\alpha_{ij}\otimes\tau^{\beta }_{kl}, \quad
c_{\mathcal U}^{\alpha\beta }:=\iint\limits_{G_1\ G_2} dg_1 dg_2
\psi_{\mathcal U}(g_1,g_2) \overline{\chi_\alpha(g_1)}
\,\overline{\chi_{\beta }(g_2)},
\end{equation}
where $\chi_\alpha(g):=\tr\pi_\alpha(g)$ is the character of the representation $\pi_\alpha$ and,
analogously, $\chi_{\beta }$ is the character of $\tau_{\beta }$.
It can be then shown that the series (\ref{fourierU}) converges uniformly for
every $\mathcal U$
(cf. Ref. \cite{Folland}, p. 137). Thus, the role of the constants
$c_{\mathcal U}^{\alpha\beta }$ is to enhance convergence of the Fourier
series (\ref{fourier}).
Note that since $f\ast\psi_{\mathcal U}=\kappa_{\mathcal V}^*\ast f \ast \kappa_{\mathcal V}$,
where the sets $\mathcal V$ are defined as in Eq. (\ref{V}) but on $\g$,
the regularization preserves positive definiteness (cf. Eq. (\ref{pdreg2})
where we proved it for measures on a single group $G_1$).
In fact, it preserves separability as well, as we
will show later (see Lemma \ref{separowalnosc}).
Finally, the initial function $f$ can be recovered from $f\ast\psi_{\mathcal U}$ by letting
$\mathcal U\to\{e_1,e_2\}$ as then $f\ast\psi_{\mathcal U}\to f$
uniformly (cf. Ref. \cite{Folland}, Theorem 2.42).

Let us define operators
$\hat f_{\alpha\beta }\in\mathcal{L}(\mh_\alpha\otimes\widetilde\mh_\beta)$
by:
\begin{equation}\label{operatorki}
\hat f_{\alpha\beta }:=\sum_{i,\dots,l}f^{jilk}_{\alpha\beta }
|e_i\rangle\langle e_j|\otimes|\tilde e_k\rangle\langle\tilde e_l|
=n_\alpha m_\beta\iint\limits_{G_1\ G_2} dg_1dg_2 f(g_1,g_2)
\pi_\alpha(g_1)^\dagger\otimes\tau_{\beta }(g_2)^\dagger,
\end{equation}
(note the change of the order of indices). Operators $\hat f_{\alpha\beta }$
are {\it inverse Fourier transforms}
of $f$ : $\hat f_{\alpha\beta }\equiv\hat f(\pi_\alpha\otimes\tau_{\beta })$
\cite{Folland,terakurwamy}
and Fourier series (\ref{fourier}) and (\ref{fourierU}) can be rewritten as:
\begin{equation}\label{tr}
f=\sum_{\alpha,\beta }\text{tr}\big[\hat f_{\alpha\beta }\,
\pi_\alpha\otimes\tau_{\beta }\big],\quad
f\ast\psi_{\mathcal U}=\sum_{\alpha,\beta }c_{\mathcal U}^{\alpha\beta }
\text{tr}\big[\hat f_{\alpha\beta }\,\pi_\alpha\otimes\tau_{\beta }\big].
\end{equation}
The last equation again explicitly shows the role of
regularization in enhancing convergence of the Fourier series (\ref{fourier}).

Having recalled the technicalities of the Fourier analysis,
we proceed to relate the group-theoretical formalism to the standard one.
We begin by quoting a standard fact, which we will extensively use
in what follows (see e.g. Ref. \cite{HewittII}, Theorem 34.10):

\begin{thm}\label{dodatniosc}
$\varphi\in\pd(\g)$ if and only if $\hat \varphi_{\alpha\beta }\geqslant 0$ for
all $[\pi_\alpha]\in\widehat G_1$ and $[\tau_{\beta }]\in\widehat G_2$.
\end{thm}

This is a non-commutative analog of the fact that positive definiteness
corresponds under (usual) Fourier transform to positivity.

We present a proof of the above theorem just for completeness' sake.
Let us first introduce an abbreviation ${\bm g}:=(g_1, g_2)\in \g$.
From Eq. (\ref{operatorki}) we then obtain for any
$v\in\mh_\alpha\otimes\widetilde\mh_\beta$:
\begin{eqnarray}
\langle v|\hat\varphi_{\alpha\beta } v\rangle &=&
n_\alpha m_\beta \iint d g_1 d g_2 \,\varphi(g_1,g_2)
\big\langle v\big|\pi_\alpha(g_1)^\dagger\otimes\tau_{\beta }(g_2)^\dagger v \big\rangle \nonumber\\
&=& n_\alpha m_\beta  \iint d {\bm h} d {\bm g} \varphi({\bm h}^{-1}{\bm g})
\big\langle \pi_\alpha(h_1)^\dagger\otimes\tau_{\beta }(h_2)^\dagger v\big|
\pi_\alpha(g_1)^\dagger\otimes\tau_{\beta }(g_2)^\dagger v\big\rangle\nonumber\\
&=& \sum_{i,k=1}^{n_\alpha,m_\beta} n_\alpha m_\beta
\iint d {\bm h} d{\bm g} \,\overline{v^{\alpha\beta }_{ik}({\bm h}})\,\varphi({\bm h}^{-1}{\bm g})
v^{\alpha\beta }_{ik}({\bm g})\geqslant 0,
\end{eqnarray}
where $v^{\alpha\beta }_{ik}({\bm h}):=\big\langle e_i\otimes
\tilde e_k \big|\pi_\alpha(h_1)^\dagger\otimes \tau_{\beta
}(h_2)^\dagger v\big\rangle$. In the second step above we inserted
$1=\iint_{\g} dh_1 dh_2$ and then changed the variables ${\bm
g}\to {\bm h}^{-1}{\bm g}$. Then we inserted the unit matrix ${\bf
1}_\alpha\otimes {\bf 1}_{\beta }$, decomposed with respect to the
fixed bases $\{e_i\}$, $\{\tilde e_k\}$  of $\mh_\alpha$,
$\widetilde\mh_\beta$ and used positive definiteness of $\varphi$.
Note that we do not need uniform convergence of the Fourier series
here and hence we used the $L^2$-convergent series (\ref{fourier})
of $\varphi$. The same applies to the proof in the other
direction.

Let us now assume that $\hat \varphi_{\alpha\beta }\geqslant 0$ for all $\alpha,\beta $.
Then from Eq. (\ref{fourier}) we obtain:
\begin{eqnarray}
& & \iint d{\bm g} d{\bm h} \overline{f({\bm g})}
\varphi({\bm g}^{-1}{\bm h})f({\bm h})=
\sum_{\alpha,\beta }\sum_{i,\dots,l}\varphi^{ijkl}_{\alpha\beta }
\iint d{\bm g} d{\bm h} \overline{f(g_1,g_2)} \pi^\alpha_{ij}(g_1^{-1}h_1)
\tau^{\beta }_{kl}(g_2^{-1}h_2)f(h_1,h_2)\nonumber\\
& & =\sum_{\alpha,\beta }\sum_{i,\dots,l, r,s}\varphi^{ijkl}_{\alpha\beta }
\iint dg_1 dg_2 \overline{f(g_1,g_2)}\:
\overline{\pi^\alpha_{ri}(g_1)}\,\overline{\tau^{\beta }_{sk}(g_2)}
 \iint dh_1 dh_2 f(h_1,h_2)\pi^\alpha_{rj}(h_1)
\tau^{\beta }_{sl}(h_2)\nonumber\\
& & =\sum_{\alpha,\beta }\sum_{r,s}
\langle c_{rs}|\hat \varphi_{\alpha\beta }\, c_{rs}\rangle\geqslant 0
\end{eqnarray}
for any $f\in C(\g)$, where $c_{rs}:=\sum_{ik}\iint dg_1 dg_2
f(g_1,g_2)\pi^\alpha_{ri}(g_1)\tau^{\beta }_{sk}(g_2) \: e_i\otimes\tilde e_k$.$\Box$
\\

Thus, by the above theorem, a positive definite function
generates a family of subnormalized states $\hat \varphi_{\alpha\beta }$.
The operators $\hat \varphi_{\alpha\beta }$ are not normalized
(except in the trivial case when sum (\ref{tr}) consists of one term only)
even if $\varphi$ is, since from Eq. (\ref{tr}) we obtain that
$\sum_{\alpha,\beta }\text{tr}(\hat \varphi_{\alpha\beta })
=\varphi(e_1,e_2)=1$,
so $\text{tr}(\hat \varphi_{\alpha\beta })\leqslant 1$.
However, we can still speak of separability of the operators
$\hat \varphi_{\alpha\beta }$ in the sense that they are decomposable into convex combinations of
products of positive operators
(cf. the corresponding remark after Definition \ref{Sep}).
The following result holds
(cf. Ref. \cite{terakurwamy} where a weaker version was proven):

\begin{lem}\label{separowalnosc}
A function $\varphi\in\pdn(\g)$ is separable if and only if the operators
$\hat\varphi_{\alpha\beta }\in\mathcal{L}(\mh_\alpha\otimes \widetilde\mh_\beta)$
are separable for all $[\pi_\alpha]\in\widehat G_1$ and
$[\tau_{\beta }]\in\widehat G_2$.
\end{lem}

Indeed, from Eq. (\ref{operatorki}) it follows that
if $Sep_0\ni\varphi=\sum_{m=1}^K p_m \e_m\otimes\eta_m$ then
$\hat \varphi_{\alpha\beta }=\sum_{m=1}^K p_m\hat\e^{(m)}_\alpha\otimes\hat\eta^{(m)}_{\beta }$,
where all the operators $\hat\e^{(m)}_\alpha, \hat\eta^{(m)}_{\beta }$
are positive by Lemma \ref{dodatniosc}, since
$\e_m\in\en(G_1), \eta_m \in\en(G_2)$ for all $m$. This extends to all of $Sep$ by the
continuity for all $\alpha, \beta $ of the inverse Fourier transform (\ref{operatorki})
$\varphi\mapsto\hat\varphi_{\alpha\beta }$
and the fact that positive separable matrices $\sigma$ with $\tr{\sigma}\leqslant 1$
form a compact convex subset of $\mathcal{L}(\mh_\alpha\otimes\widetilde\mh_\beta)$
\cite{Pawel}.
The latter follows from the facts that i) the set of extreme points of the latter subset
can be identified with
$\mathbb{C}P^{n_\alpha}\times\mathbb{C}P^{m_\beta}\cup\{0\}$
and ii) the convex hull of a compact subset of $\mathbb R^N$ is compact.

Conversely, assume that for every $\alpha,\beta $, $\hat
\varphi_{\alpha\beta } =\sum_{m=1}^{K_{\alpha\beta
}}p_m^{\alpha\beta }\,|x_m^\alpha\rangle\langle x_m^\alpha|
\otimes |y_m^{\beta }\rangle\langle y_m^{\beta }|$, where
$x_m^\alpha\in\mh_\alpha$, $y_m^{\beta }\in\widetilde\mh_\beta$.
We will prove that $\varphi$ is then a uniform limit of separable
functions and hence is itself separable. First, we pass to the
regularized function $\varphi_{\mathcal
U}:=\varphi\ast\psi_{\mathcal U}$, according to the procedure we
described above (cf. Eq. (\ref{fourierU}) and the surrounding
paragraph). As we mentioned, $\varphi_{\mathcal U}$ is positive
definite and hence $\varphi_{\mathcal
U}(e_1,e_2)=||\varphi_{\mathcal U}||_\infty> 0$ (except in the
trivial case $\varphi\equiv 0$), which allows us to pass to the
normalized function $\varphi_{\mathcal U}/||\varphi_{\mathcal
U}||_\infty$. Then Eq. (\ref{tr}) implies that:
\begin{eqnarray}
& &\frac{1}{||\varphi_{\mathcal U}||_\infty}\varphi_{\mathcal U}(g_1,g_2)
=\frac{1}{||\varphi_{\mathcal U}||_\infty}\sum_{\alpha,\beta }
\sum_{m=1}^{K_{\alpha\beta }}c_{\mathcal U}^{\alpha\beta }p_m^{\alpha\beta }\,
\langle x_m^\alpha| \pi_\alpha(g_1)x_m^\alpha\rangle
\langle y_m^{\beta }| \tau_{\beta }(g_2) y_m^{\beta }\rangle\nonumber\\
& &= \sum_{\alpha,\beta }\sum_{m=1}^{K_{\alpha\beta }}\frac{1}{||\varphi_{\mathcal U}||_\infty}
c_{\mathcal U}^{\alpha\beta }p_m^{\alpha\beta }||x_m^\alpha||^2||y_m^{\beta }||^2\,
\Big\langle \frac{x_m^\alpha}{||x_m^\alpha||}\Big|
\pi_\alpha(g_1)\frac{x_m^\alpha}{||x_m^\alpha||}\Big\rangle
\Big\langle \frac{y_m^{\beta }}{||y_m^{\beta }||}\Big| \tau_{\beta }(g_2)
\frac{y_m^{\beta }}{||y_m^{\beta }||}\Big\rangle\label{cipa}
\end{eqnarray}
and the series converges uniformly. The functions given by scalar products
belong to $\en(G_1)$ and $\en(G_2)$ respectively, since
$\pi_\alpha$ and $\tau_{\beta }$ are irreducible.
From their definition in Eq. (\ref{fourierU}) and the definition
of $\psi_{\mathcal U}$ (\ref{reg}) it also follows that
the factors $c_{\mathcal U}^{\alpha\beta }$ are non-negative:
\begin{eqnarray}
c_{\mathcal U}^{\alpha\beta }&=&\iint d {\bm g} d {\bm h} \kappa_{\mathcal V}({\bm h})
\kappa_{\mathcal V}({\bm h}^{-1}{\bm g})\overline{\chi_{\alpha\beta }({\bm g})}
=\iint d {\bm g} d {\bm h} \kappa_{\mathcal V}({\bm h}^{-1})
\kappa_{\mathcal V}({\bm g})\overline{\chi_{\alpha\beta }({\bm h}{\bm g})}\nonumber\\
&=&\iint d {\bm g} d {\bm h} \kappa_{\mathcal V}({\bm h})
\kappa_{\mathcal V}({\bm g})\overline{\chi_{\alpha\beta }({\bm h}^{-1}{\bm g})}
= \frac{1}{n_\alpha^2 m_\beta^2}
\overline{\tr\big[\widehat{\kappa_{\mathcal V}}_{\alpha\beta}
(\widehat{\kappa_{\mathcal V}}_{\alpha\beta})^\dagger\big]}
\geqslant 0,\label{cu>}
\end{eqnarray}
where we used definition (\ref{operatorki}) and the fact that
$\kappa_{\mathcal V}$ is symmetric (cf. property $(i)$ in Eq. (\ref{V})) and real.
Evaluating $\varphi_{\mathcal U}/||\varphi_{\mathcal U}||_\infty$ at the neutral element
we see that the sum in Eq. (\ref{cipa}) is in fact a convex combination
of pure product functions (cf. the definition of a pure function in Section 2), since:
\begin{equation}
\sum_{\alpha,\beta }\sum_{m=1}^{K_{\alpha\beta }}\frac{1}{||\varphi_{\mathcal U}||_\infty}
c_{\mathcal U}^{\alpha\beta }p_m^{\alpha\beta }
||x_m^\alpha||^2||y_m^{\beta }||^2=
\frac{\varphi_{\mathcal U}(e_1,e_2)}{||\varphi_{\mathcal U}||_\infty}=1
\ \ \text{and}\ \ \frac{1}{||\varphi_{\mathcal U}||_\infty}
c_{\mathcal U}^{\alpha\beta } p_m^{\alpha\beta }
||x_m^\alpha||^2||y_m^{\beta }||^2\geqslant 0.
\end{equation}
Thus, $\varphi_{\mathcal U}/||\varphi_{\mathcal U}||_\infty$ is separable, as a uniform limit of
separable functions. Since $\varphi_{\mathcal U}
\xrightarrow[\mathcal U\to\{e_1,e_2\}]{} \varphi$
uniformly, $\varphi\in Sep$. $\Box$
\\

From Lemma \ref{separowalnosc} it follows that the problem of describing
separable functions on $\g$ generates a family of separability
problems in all pairs of dimensions where $G_1$ and $G_2$ have
irreducible representations. In other words, it plays a role of a
``generating function'' for this family. Conversely, from the form
of the Fourier transformation (\ref{glowny}) and its inverse
(\ref{operatorki}) it follows that density matrices on
$\mh_\alpha\otimes\widetilde\mh_\beta$ are in one-to-one
correspondence with those functions $\varphi$ from $\pdn(\g)$,
which belong to the (finite-dimensional) linear span of
$\pi^{\alpha}_{ij}\otimes\tau^{\beta }_{kl}$, where
$\pi_\alpha,\tau_{\beta }$ are fixed. Moreover, since for an
arbitrary density matrix
$\varrho\in\mathcal{L}(\mh_\alpha\otimes\widetilde\mh_\beta)$ its
Fourier transform $\varphi_\varrho$ (cf. Eq. (\ref{glowny}))
satisfies:
\begin{equation}\label{cos}
(\widehat{\varphi_\varrho})_{\gamma\nu }=\delta_{\alpha\gamma}\delta_{\beta \nu }\varrho,
\end{equation}
Lemma \ref{separowalnosc} implies that (cf. Ref. \cite{terakurwamy}, Theorem 1):

\begin{cor}\label{separ}
A state $\varrho$ is separable if and only if
$\varphi_\varrho\in Sep$ .
\end{cor}

Next we examine bounded linear maps $\Lambda\colon C(G_2)\to C(G_1)$.
For an arbitrary  function $f\in C(G_2)$, we consider a function
$\Lambda f_U\in C(G_1)$, where $f_U:=f\ast \psi_U$ is the regularization of $f$.
Now, the regularizing functions $\psi_U\in C(G_2)$ are the
single-group functions defined in Eq. (\ref{reg}), but now on the group $G_2$, and
the sets $U$ run through a neighborhood base of $e_2\in G_2$.
Note, however, that unlike in Section \ref{GrHor} here we are regularizing
the argument of $\Lambda$ and not its value.
Calculating Fourier transform of $\Lambda f_U$ from the single-group version
of the definition (\ref{operatorki}) we obtain:
\begin{eqnarray}
\big(\widehat{\Lambda f_U}\big)_{\alpha}&=&n_{\alpha}\int\limits_{G_1} dg_1 \Lambda\big(f\ast\psi_U\big)(g_1)
\pi_{\alpha}(g_1)^\dagger=n_{\alpha}\int\limits_{G_1} dg_1 \sum_{\beta } c_U^{\beta }\sum_{k,l}f^{kl}_{\beta }
\big(\Lambda \tau_{kl}^{\beta }\big)(g_1)\pi_{\alpha}(g_1)^\dagger\nonumber\\
&=&\sum_{\beta } c_U^{\beta }\sum_{k,l}f^{kl}_{\beta }\,
n_{\alpha}\int\limits_{G_1} dg_1 \big(\Lambda\tau_{kl}^{\beta } \big)(g_1)\pi_{\alpha}(g_1)^\dagger,\label{chuj}
\end{eqnarray}
where we used the uniform convergence of
Fourier series for $f_U$
and the fact that $\Lambda$ is continuous in the uniform norm. The regularizing
constants $c^{\beta }_U$ are defined analogously as in Eq. (\ref{fourierU}), i.e.
\begin{equation}
c^{\beta }_U:=\int\limits_{G_2} dg_2 \psi_U(g_2) \overline{\chi_{\beta }(g_2)}.
\end{equation}
We will find it useful to define maps $\hat\Lambda_{\alpha}^{\beta }\colon
\mathcal{L}(\widetilde\mh_\beta)\to\mathcal{L}(\mh_\alpha)$ through
an analog of Eq. (\ref{operatorki}):
\begin{equation}\label{Lop}
\hat\Lambda_{\alpha}^{\beta }:=\sum_{i,\dots,l}\Lambda^{ji\beta }_{\alpha lk}
|E_{ij}\rangle_{\scriptscriptstyle{HS}} \langle\widetilde E_{kl} |,\quad
\Lambda^{ij\beta }_{\alpha kl}:=n_\alpha \int\limits_{G_1} dg_1
\,\overline{\pi^\alpha_{ij}(g_1)}\big(\Lambda \tau^{\beta }_{kl}\big)(g_1)
\end{equation}
(note the change of the order of indices),
where $\widetilde E_{kl}:=|\tilde e_k\rangle\langle \tilde e_l|$,
$E_{ij}:=|e_i\rangle\langle e_j|$
are the bases of $\mathcal{L}(\widetilde\mh_\beta)$ and $\mathcal{L}(\mh_{\alpha})$ respectively,
and $\langle A|B\rangle_{\scriptscriptstyle{HS}}=\text{tr}(A^\dagger B)$.
We can then rewrite Eq. (\ref{chuj}) as follows:
\begin{equation}\label{chuj2}
\big(\widehat{\Lambda f_U}\big)_\alpha=\sum_{\beta } c_U^{\beta }
\hat\Lambda_{\alpha}^{\beta } \hat f_{\beta }.
\end{equation}
Note that in the above series all operators
$\hat\Lambda_{\alpha}^{\beta } \hat f_{\beta }\in\mathcal{L}(\mh_\alpha)$
are finite-dimensional and hence the convergence can be understood
in any of the equivalent norms on $\mathcal{L}(\mh_\alpha)$.

Conversely, given an arbitrary map
$\Phi\colon\mathcal{L}(\widetilde\mh_\beta)\to\mathcal{L}(\mh_\alpha)$
we can Fourier transform it and assign to it a map $\Lambda_\Phi\colon C(G_2)\to C(G_1)$
through the following formula
(compare with Eq. (\ref{glowny}) where Fourier transform of operators was defined):
\begin{equation}\label{FA}
\Lambda_\Phi f(g_1):=\tr\Big[\big(\Phi\hat f_{\beta }\big)\pi_{\alpha}(g_1)\Big]
\quad \text{for every}\ f\in C(G_2).
\end{equation}
It is obvious that $\Lambda_\Phi f\in C(G_1)$, because we consider only
continuous representations.
Moreover from the definition of $\hat f_{\beta }$ we have:
\begin{eqnarray}
||\Lambda_\Phi f||_\infty&=&m_\beta \sup_{g_1\in G_1}
\Bigg|\int\limits_{G_2} dg_2 f(g_2)
\tr\Big[\big(\Phi\tau_{\beta }^\dagger(g_2)\big)\pi_{\alpha}(g_1)\Big]\Bigg|\nonumber\\
&\leqslant& m_\beta ||f||_\infty \sup_{g_1\in G_1}\int\limits_{G_2} dg_2\bigg|
\tr\Big[\big(\Phi\tau_{\beta }^\dagger(g_2)\big)\pi_{\alpha}(g_1)\Big]\bigg|.
\end{eqnarray}
The last supremum is finite, as the integrand is continuous
and $G$ is compact, and independent of $f$,
so that $\Lambda_\Phi$ is bounded. The transformation (\ref{FA}) is an inverse of
the mapping $\Lambda \mapsto \hat\Lambda_{\alpha}^{\beta }$ given by Eq. (\ref{Lop}),
since by an easy direct calculation one finds that (compare Eq. (\ref{cos})):
\begin{equation}\label{cos2}
(\widehat\Lambda_\Phi)^{\mu }_{\nu}=\delta_{\nu\alpha}\delta_{\mu \beta } \Phi.
\end{equation}


By analogy with Theorem \ref{dodatniosc}, which
characterizes positive definite functions in terms of their
inverse Fourier transforms, one would expect a corresponding
characterization of positive definite (PD) maps $\Lambda$ (cf. Definition
\ref{posdef}) in terms of their inverse
Fourier transforms $\hat\Lambda^{\beta }_{\alpha}$.
The next lemma provides such
a characterization:

\begin{lem}\label{dodatniosc2}
A bounded linear map $\Lambda\colon C(G_2)\to C(G_1)$ is positive
definite, i.e. $\Lambda \pd(G_2)\subset\pd(G_1)$, if and only if
the maps $\hat\Lambda^{\beta }_{\alpha}\colon
\mathcal{L}(\widetilde\mh_\beta)\to\mathcal{L}(\mh_\alpha)$ are
positive for all $[\pi_\alpha]\in\widehat G_1,[\tau_{\beta
}]\in\widehat G_2$.
\end{lem}

To prove it, let us take an arbitrary $\phi\in\pd(G_2)$, which by
Theorem \ref{dodatniosc} is equivalent to $\hat\phi_{\beta
}\geqslant 0$ for all $\beta $. We employ Eq. (\ref{chuj2}). If
all maps $\hat\Lambda^{\beta }_{\alpha}$ are positive, then
$c_U^{\beta }\hat\Lambda_{\alpha}^{\beta } \hat \phi_{\beta
}\geqslant 0$ for all $\beta $, since $c_U^{\beta }\geqslant 0$
(cf. Eq. (\ref{cu>}) where we proved it for $\g$). Since positive
operators form a closed cone in $\mathcal{L}(\mh_\alpha)$, the
series $\sum_{\beta } c_U^{\beta }\hat\Lambda_{\alpha}^{\beta }
\hat \phi_{\beta }$ converges to a positive operator and hence
$\big(\widehat{\Lambda \phi_U}\big)_\alpha\geqslant 0$ for all
$\alpha$. Theorem \ref{dodatniosc} implies then that
$\Lambda(\phi\ast\psi_U)\in\pd(G_1)$. Taking the limit
$U\to\{e_2\}$, $\phi\ast\psi_U$ converges to $\phi$ uniformly.
Thus, using continuity of $\Lambda$ and uniform closedness of
$\pd(G_1)\subset C(G_1)$, we obtain that $\Lambda\phi\in\pd(G_1)$
for any $\phi\in\pd(G_2)$.

Conversely, assume $\Lambda$ to be positive definite. Let us
fix $[\tau_{\beta }]\in\widehat G_2$ and take an arbitrary positive operator
$\varrho\in\mathcal{L}(\widetilde\mh_\beta)$. Applying Fourier transform
(\ref{glowny}) to $\varrho$ we obtain its characteristic function:
\begin{equation}
\phi_\varrho(g_2):=\text{tr}\big[\varrho\tau_{\beta }(g_2)\big]
=\sum_{k,l}\varrho^{kl}\tau_{lk}^{\beta }(g_2),
\end{equation}
which by Theorem \ref{dodatniosc} is positive definite, since
$(\widehat{\phi_\varrho})_{\gamma }=\delta_{\gamma \beta }\varrho\geqslant 0$.
Hence, $\Lambda \phi_\varrho$ is positive definite too. Then from
Eq. (\ref{chuj2}), where we can neglect the regularization and put
$c_U^{\beta }=1$ since the Fourier series of $\phi_\varrho$ contains only
one term, and from Theorem \ref{dodatniosc} we obtain that
$\big(\widehat{\Lambda \phi_\varrho}\big)_\alpha=
\hat\Lambda_{\alpha}^{\beta }\varrho\geqslant 0$ for all $\alpha$.
Since $\tau_{\beta }$ and $\varrho$ were arbitrary, the result follows.$\Box$
\\

Thus, from the above lemma and Eq. (\ref{cos2}) it follows that
(compare Corollary \ref{separ}):

\begin{cor}
A map $\Phi\colon\mathcal{L}(\widetilde\mh_\beta)\to\mathcal{L}(\mh_\alpha)$
is positive if and only if the map $\Lambda_\Phi\colon C(G_2)\to C(G_1)$,
defined in Eq. (\ref{FA}), is positive definite.
\end{cor}

Next we present a characterization of completely positive definite maps
(cf. Definition \ref{posdef}) in terms of their
Fourier transforms. We first prove the following fact:

\begin{lem}\label{cpd}
A bounded linear map $\Lambda\colon C(G_2)\to C(G_1)$ is $G_2$-positive definite,
i.e. $(\I\otimes\Lambda) \pd(G_2\times G_2)\subset\pd(G_2\times G_1)$, if and only if
maps $\hat\Lambda^{\beta }_{\alpha}\colon
\mathcal{L}(\widetilde\mh_\beta)\to\mathcal{L}(\mh_\alpha)$
are completely positive for all
$[\pi_\alpha]\in\widehat G_1,[\tau_{\beta }]\in\widehat G_2$.
\end{lem}

Let $\varphi\in\pd(G_2\times G_2)$, so by Theorem \ref{dodatniosc}
$\hat\varphi_{\beta \gamma }\geqslant 0$ for all
$[\tau_\beta],[\tau_\gamma]\in \widehat G_2$.
Then the obvious generalization of Eq. (\ref{chuj2}) to $G_2\times G_2$,
implies that:
\begin{equation}\label{chuj3}
\Big[\widehat{(\I\otimes\Lambda)\varphi_{\mathcal U}}\Big]_{\beta \alpha}
=\sum_{\delta ,\gamma } c_{\mathcal U}^{\delta \gamma }\big(\widehat{\I\otimes
\Lambda}\big)^{\delta \gamma }_{\beta \alpha}\,\hat\varphi_{\delta \gamma }
=\sum_{\gamma } c_{\mathcal U}^{\beta \gamma }({\bm 1}_{\beta }\otimes
\hat\Lambda^{\gamma }_{\alpha})\hat\varphi_{\beta \gamma },
\end{equation}
where now $\mathcal U\subset G_2\times G_2 $ runs through a
neighborhood base of $\{e_2,e_2\}\in G_2\times G_2 $.
If all maps $\hat\Lambda_{\alpha}^{\gamma }$ are completely positive, then
$c_{\mathcal U}^{\beta \gamma }({\bm 1}_{\beta }\otimes
\hat\Lambda^{\gamma }_{\alpha})\hat\varphi_{\beta \gamma }\geqslant 0$
as operators from $\mathcal{L}(\widetilde\mh_\beta\otimes\widetilde\mh_\alpha)$,
since $c_{\mathcal U}^{\beta \gamma }\geqslant 0$ for all
$\beta , \gamma $ (cf. Eq. (\ref{cu>})).
From closedness of the cone of positive
operators in
$\mathcal{L}(\widetilde\mh_\beta\otimes\widetilde\mh_\alpha)$, the
series in Eq. (\ref{chuj3}) converges to a positive operator as
well. Hence $\Big[\widehat{(\I\otimes\Lambda)\varphi_{\mathcal
U}}\Big]_{\beta \alpha}\geqslant 0$ for all $\alpha, \beta $ and
from Theorem \ref{dodatniosc} it follows that
$(\I\otimes\Lambda)(\varphi\ast\psi_{\mathcal U})\in\pd(G_2\times
G_1)$. We remove the regularization by letting $\mathcal
U\to\{e_2,e_2\}$ so that $\varphi\ast\psi_{\mathcal U}\to\varphi$
uniformly. Then from the continuity of $\I\otimes\Lambda$ and
uniform closedness of $\pd(G_2\times G_1)$ it follows that
$(\I\otimes\Lambda)\varphi\in\pd(G_2\times G_1)$ for any
$\varphi\in\pd(G_2\times G_2)$.

For the proof in the other direction, we proceed along the same
lines as in the proof of the previous lemma---for arbitrary
$[\tau_{\beta }],[\tau_{\gamma }] \in \widehat G_2$ and arbitrary
$0\leqslant
\varrho\in\mathcal{L}(\widetilde\mh_\beta\otimes\widetilde\mh_{\gamma
})$, we consider the Fourier transform of $\varrho$,
$\varphi_\varrho$, defined in Eq. (\ref{glowny}). Then positive
definiteness of $\I\otimes\Lambda$, Theorem \ref{dodatniosc}, and
Eq.(\ref{chuj2}) generalized to $G_2\times G_2$ (with $c^{\mu \nu
}_{\mathcal U}=1$ as the Fourier series (\ref{fourier}) of
$\varphi_\varrho$ contains only one term) imply that
$\Big[\widehat{(\I\otimes\Lambda)\varphi_\varrho}\Big]_{\beta
\alpha}= ({\bm 1}_{\beta }\otimes\hat\Lambda^{\gamma
}_{\alpha})\varrho\geqslant 0$ for all $\gamma $. Since
$[\tau_{\beta }]$, $[\tau_{\gamma }]$, and $\varrho$ are
arbitrary, $G_2$-positive definiteness of $\Lambda$ implies that
every map $\hat\Lambda^{\gamma }_{\alpha}$ is
$(\text{dim}\tau)$-positive for all possible $[\tau]\in\widehat
G_2$. Thus, in particular, every $\hat\Lambda^{\gamma }_{\alpha}$
is $m_{\gamma }$-positive, $m_{\gamma
}=\text{dim}\widetilde\mh_{\gamma }$. But then by the Choi Theorem
(cf. Ref. \cite{Choi}, Theorem 2) this is equivalent to
$\hat\Lambda^{\gamma }_{\alpha}$ being completely positive.$\Box$
\\

As a by-product we obtain an analog of the Choi Theorem
(cf. Ref. \cite{Choi}, Theorem 2) for PD maps:

\begin{cor}\label{Choi}
A map $\Lambda\colon C(G_2)\to C(G_1)$ is completely positive definite if and only if
it is $G_2$-positive definite.
\end{cor}

The proof in one direction follows immediately from Definition
\ref{posdef}. For the opposite implication, let us assume that
$\Lambda$ is $G_2$-PD. Let $H$ be an arbitrary compact group and
let Latin indices $a,b,\dots$ enumerate irreps of $H$.
From the complete positivity of Fourier transforms
$\hat\Lambda_{\alpha}^{\beta }$, guaranteed by Lemma
\ref{cpd}, Theorem \ref{dodatniosc}, Eq. (\ref{cu>}) applied to $H\times G_2$, and closedness of the cone of
positive operators, we obtain that:
\begin{equation}
\Big[\widehat{(\I\otimes\Lambda)\varphi_{\mathcal U}}\Big]_{b\alpha}
=\sum_{\beta } c_{\mathcal U}^{b\beta }({\bm 1}_b\otimes
\hat\Lambda^{\beta }_{\alpha})\hat\varphi_{b\beta }\geqslant 0,
\end{equation}
for every $b$ and $\alpha$ and every $\varphi\in\pd(H\times G_2)$.
Thus from Theorem \ref{dodatniosc}, $(\I\otimes\Lambda)
\varphi_{\mathcal U}\in\pd(H\times G_1)$.
Letting $\mathcal U\to \{e_H, e_2\}\in H\times G_2$, so that
$\varphi_{\mathcal U}$ converges to $\varphi$ uniformly,
and using continuity of
$\I\otimes\Lambda$ and uniform closedness of $\pd(H\times G_1)$,
we obtain that $(\I\otimes\Lambda)
\varphi\in\pd(H\times G_1)$ for every compact $H$ and every
$\varphi\in\pd(H\times G_2)$. $\Box$
\\

From Lemma \ref{cpd} and Corollary \ref{Choi} we finally obtain:

\begin{lem}\label{Choi2}
A bounded linear map $\Lambda\colon C(G_2)\to C(G_1)$ is
completely positive definite if and only if the maps
$\hat\Lambda^{\beta }_{\alpha}\colon
\mathcal{L}(\widetilde\mh_\beta)\to\mathcal{L}(\mh_\alpha)$ are
completely positive for all $[\pi_\alpha]\in\widehat
G_1,[\tau_{\beta }]\in\widehat G_2$.
\end{lem}

Comparison of Theorem \ref{main} with the Horodecki Theorem
\ref{hor} shows that positive definite mappings of continuous
functions on compact groups play an analogous role to that of
positive mappings of density matrices in the standard theory of
entanglement \cite{Horodeccy}. The harmonic-analytical formalism
described above makes this observation, as well as the
``generating function'' analogy, formal. Namely, Lemma
\ref{dodatniosc2} implies that every PD map $\Lambda$ generates a
family of positive maps $\hat\Lambda^{\beta }_{\alpha}$, acting
between algebras of operators on representation spaces of $G_1$
and $G_2$. Conversely, Fourier transform (\ref{FA}), together with
property (\ref{cos2}) and Lemma \ref{dodatniosc2} allows one to
assign a unique PD map to every suitable (cf. definition
(\ref{FA})) positive map. Analogously, Lemma \ref{Choi2} shows
that each CPD map gives rise to a family of completely positive
maps, and by Fourier transform (\ref{FA}) every (suitable)
completely positive map defines a CPD map. Thus, PD and CPD maps
between groups play a role of ``generating functions'' of families
of positive and completely positive maps respectively.

In order to compare Theorem \ref{main} with the Horodecki Theorem
\ref{hor}, we first choose $G_1$ and $G_2$ so that they possess
irreps in dimensions $\text{dim}\mh_{\mathscr A}$ and
$\text{dim}\mh_{\mathscr B}$, so that we may identify
$\mh_{\mathscr A}\cong\mh_\alpha$ and $\mh_{\mathscr
B}\cong\widetilde\mh_\beta$ for some $[\pi_\alpha]\in\widehat G_1$
and $[\tau_{\beta }]\in\widehat G_2$ \cite{terakurwamy}. Apart
from that there are no further restrictions on $G_1,G_2$. Finding
such a group for a given finite-dimensional $\mh_{\mathscr
A},\mh_{\mathscr B}$ is always possible---for example we can take
$G_1=G_2=SU(2)$, which possesses irreps in all finite dimensions.
Having made the above identification, we obtain from Theorem
\ref{main} that:

\begin{thm}\label{grouphor}
A density matrix $\varrho\in\mathcal{L}(\mh_\alpha\otimes\widetilde\mh_\beta)$ is separable
if and only if for all $[\pi_\gamma]\in \widehat G_1$ and all positive
maps $\Phi_\gamma\colon\mathcal{L}(\widetilde\mh_\beta)\to\mathcal{L}(\mh_\gamma)$,
$({\bm 1}_\alpha\otimes\Phi_\gamma)\varrho\geqslant 0$ as an operator on
$\mh_\alpha\otimes\mh_\gamma$.
\end{thm}

To prove it, we first Fourier transform $\varrho$,
passing to its characteristic function
$\varphi_\varrho=\text{tr}(\varrho\pi_{\alpha}\otimes\tau_{\beta })\in\pdn(\g)$.
From Lemma \ref{separ} $\varrho$ is separable if and only if $\varphi\in Sep$.
Applying Theorem \ref{main} to $\varphi_\varrho$ we obtain
that $\varrho$ is separable if and only if for all positive definite maps
$\Lambda\colon C(G_2)\to C(G_1)$,
$(\I\otimes\Lambda)\varphi_\varrho\in\pd(G_1\times G_1)$. From the generalization of Eq. (\ref{chuj2})
to $\g$ with all $c_{\mathcal U}^{\delta\gamma }=1$ (no regularization of
$\varphi_\varrho$ is needed because the Fourier series of $\varphi_\varrho$ contains
only one non-zero term; cf. Eq. (\ref{cos}))
we obtain that:
\begin{equation}\label{ixphi}
\Big[\widehat{(\I\otimes\Lambda)\varphi_\varrho}\Big]_{\alpha\gamma}=
({\bm 1}_\alpha\otimes\hat\Lambda^{\beta }_{\gamma})\varrho.
\end{equation}
Then from Theorem \ref{dodatniosc} and Lemma \ref{dodatniosc2}
it follows that $(\I\otimes\Lambda)\varphi_\varrho$ is a positive definite function
if and only if
$({\bm 1}_\alpha\otimes\hat\Lambda^{\beta }_{\gamma})\varrho\geqslant 0$ for every $\gamma$, where
every map $\hat\Lambda^{\beta }_{\gamma}\colon
\mathcal{L}(\widetilde\mh_\beta)\to\mathcal{L}(\mh_\gamma)$ is positive.$\Box$
\\

Thus, when applied to finite dimension, Theorem \ref{main} turns out to be
weaker then Theorem \ref{hor}, since generically one has to check
positive maps operating between the fixed space $\mathcal{L}(\widetilde\mh_\beta)$
and the {\it whole family} of spaces $\mathcal{L}(\mh_\gamma)$, $[\pi_\gamma]\in\widehat G_1$,
and not only between $\mathcal{L}(\widetilde\mh_\beta)$ and $\mathcal{L}(\mh_\alpha)$
as in Theorem \ref{hor}. Note, however, that
the number of spaces $\mh_\gamma$ to check need not be infinite, since it may be possible
to find discrete $G_1$ \cite{terakurwamy}.
This is particularly easy for low-dimensional initial spaces $\mh_{\mathscr A}$,
 $\mh_{\mathscr B}$.

\section{Examples of positive definite maps for $G_1=G_2$}\label{examples}
From the point of view of classification of
separable functions using Theorem \ref{main}
only those positive definite maps which are
not completely positive definite
(cf. Definition \ref{posdef})
are interesting: if $\Lambda$ is a CPD map then all the
functions $(\I\otimes\Lambda)\varphi$, $\varphi\in\pdn(\g)$,
are positive definite,
whether $\varphi$ is separable or not.
Hence we encounter a similar problem as in the finite-dimensional
linear algebra \cite{Woronowicz}:  classify all positive definite
but not completely positive definite maps from $C(G_2)$ to $C(G_1)$.
In this Section we give some examples of PD and CPD maps for the case
$G_1=G_2\equiv G$.

The first example of PD but not CPD map
was already encountered
in Theorem \ref{PPT}---the inversion map $\theta$:
\begin{equation}\label{inv}
\theta f(g):=f(g^{-1}).
\end{equation}
To show that $\theta$ is not CPD (positive definiteness will be proven
below in a more general setting), observe that $\theta$ corresponds through
the Fourier transform to the transposition map $T$,
$T\varrho:=\varrho^T$, acting on each representation space $\mh_\alpha$
of $G$:
\begin{equation}\label{pizda}
f(g^{-1})=\sum_\alpha\sum_{i,j}f^{ij}_\alpha\pi_{ij}^\alpha(g^{-1})
=\sum_\alpha\sum_{i,j}f^{ji}_\alpha\,\overline{\pi_{ij}^\alpha(g)},
\quad \text{and hence}\quad
\big(\widehat{\theta f}\big)_{\bar \alpha}=T\hat f_\alpha=\hat f_\alpha^T.
\end{equation}
Here index $\bar \alpha$
denotes the complex conjugate $\overline{\pi}_\alpha$ of representation $\pi_\alpha$:
$\overline{\pi}_\alpha(g):=\overline{\pi_\alpha(g)}$.
Eq. (\ref{pizda}) establishes the connection between the PPT criterion
(Theorem \ref{ppt}) and Theorem \ref{PPT} (see Ref. \cite{terakurwamy} for more
details).
Now, let $\varrho\in\mathcal{L}(\mh_\alpha\otimes\mh_\beta)$ be any positive
operator such that its partial transpose
$({\bm 1}_\alpha\otimes T)\varrho$ is not positive.
Then, by Theorem \ref{dodatniosc},
the Fourier transform $\varphi_\varrho$ of $\varrho$ is positive definite,
but $(\I\otimes\theta)\varphi_\varrho$ is not. We propose to use the same
terminology as in quantum information theory and
call entangled functions $\varphi$ not detected by $\theta$
(see the remark after Theorem \ref{main})
{\it bound entangled}.

Let us consider more general substitutions of the argument in the
tested function. Let ${\bm \alpha}$ be an arbitrary automorphism
of $G$ and ${\bm \beta}$ an arbitrary anti-automorphism of $G$,
i.e.:
\begin{eqnarray}
{\bm \alpha}(gh)&=&{\bm \alpha}(g){\bm \alpha}(h),\label{hom}\\
{\bm \beta}(gh)&=&{\bm \beta}(h){\bm \beta}(g).\label{antihom}
\end{eqnarray}
We define the corresponding maps from $C(G)$ to $C(G)$:
\begin{equation}
\Lambda_{\bm \alpha} f(g):=f({\bm \alpha}(g)),\quad
\Lambda_{\bm \beta} f(g):=f({\bm \beta}(g))\label{hommap}.
\end{equation}
Both $\Lambda_{\bm \alpha}$ and $\Lambda_{\bm \beta}$ are positive definite,
which follows most directly from the GNS construction (cf. Theorem \ref {GNS}):
\begin{eqnarray}
\iint dgdh \overline{f(g)}\big(\Lambda_{\bm \alpha}\phi\big)(g^{-1}h)f(h)
&=&\Big\langle\int dg f(g) \pi_\phi\big({\bm \alpha}(g)\big)v_\phi\Big|
\int dh f(h) \pi_\phi\big({\bm \alpha}(h)\big)v_\phi\Big\rangle\geqslant 0,\nonumber\\
\iint dgdh \overline{f(g)}\big(\Lambda_{\bm \beta}\phi\big)(g^{-1}h)f(h)
&=&\Big\langle\int dh \overline{f(h)} \pi_\phi\big({\bm \beta}(h)\big)^\dagger v_\phi\Big|
\int dg \overline{f(g)} \pi_\phi\big({\bm \beta}(h)\big)^\dagger v_\phi\Big\rangle\geqslant 0,\nonumber
\end{eqnarray}
where $\phi\in\pd(G)$ and we used the fact that ${\bm
\alpha}(g^{-1})={\bm \alpha}(g)^{-1}$ and ${\bm
\beta}(g^{-1})={\bm \beta}(g)^{-1}$. Moreover, maps arising from
automorphisms are completely positive definite. Indeed, from
Corollary \ref{Choi} it is enough to check the extension of
$\Lambda$ to $C(G\times G)$. But then we obtain that (with the
boldface characters denoting elements of $G \times G$):
\begin{eqnarray}
& & \iint d{\bm g} d{\bm h} \overline{f({\bm g})}\big(\I\otimes\Lambda_{\bm \alpha}\varphi\big)
({\bm g}^{-1}{\bm h})f({\bm h})=\nonumber \\
& & \Big\langle \iint dg_1dg_2 f(g_1,g_2)
\pi_\varphi\big(g_1,{\bm \alpha}(g_2)\big)v_\varphi\Big|\iint dh_1dh_2 f(h_1,h_2)
\pi_\varphi\big(h_1,{\bm \alpha}(h_2)\big)v_\varphi\Big\rangle\geqslant 0,\label{IxLA}
\end{eqnarray}
for an arbitrary $f\in C(G\times G)$. The maps arising from
anti-automorphisms are not necessarily CPD---the above calculation
leading to the inequality (\ref{IxLA}) cannot be repeated.
However, since every anti-automorphism can be written in the form
${\bm \beta}(g)=\big[{\bm \alpha}(g)\big]^{-1}$, where ${\bm
\alpha}$ is an automorphism, every map $\Lambda_{\bm \beta}$
arising from an anti-automorphism is of the form:
\begin{equation}\label{ICPD}
\Lambda_{\bm \beta}=\Lambda^{CPD}\circ \theta,
\end{equation}
where $\Lambda^{CPD}$ is some CPD map. Hence, the use of a general
anti-homomorphism ${\bm \beta}$ in Theorem \ref{PPT} gives no
improvement, since the function $(\I\otimes\Lambda_{\bm
\beta})\varphi$ is positive definite if $(\I\otimes
\theta)\varphi$ is. In other words, PD maps of the type
(\ref{ICPD}) cannot detect bound entangled functions. This is in
close analogy to what one encounters in the study of standard
separability problems. Indeed, from Lemma \ref{cpd}, Corollary
\ref{Choi}, and Eq. (\ref{pizda}) PD maps of the form (\ref{ICPD})
generate positive maps of the type $\Phi^{CP}\circ T$, where
$\Phi^{CP}$ is a completely positive map. Clearly, by Theorem
\ref{hor}, such maps cannot detect bound entangled states
$\varrho$, for which $({\bm 1}\otimes T)\varrho\geqslant 0$.

We can give a more general example of a CPD map, motivated by
the Kraus decomposition of a completely positive map
(cf. Ref. \cite{Kraus} and Ref. \cite{Choi}, Theorem 1). For
an arbitrary measure $\mu$ from $M(G)$ we define a map:
\begin{equation}\label{Lmu}
\Lambda_\mu f(g):= \big(\mu^*\ast f \ast\mu\big)(g)=
\iint\limits_{G\ G} d\overline{\mu(a)}d\mu(b)f(agb^{-1}),
\end{equation}
where the adjoint $\mu^*$ is defined as $\mu^*(\Omega):=\overline{\mu(\Omega^{-1})}$
for any Borel set $\Omega\subset G$
(cf. the corresponding definition for functions after Eq. (\ref{pdm}))
and the convolution is defined through Eq. (\ref{splot1}).
Obviously, $\Lambda_\mu$ maps $C(G)$ to $C(G)$ and
is a generalization of the regularization
formula (\ref{pdreg}). The map $\Lambda_\mu$ is bounded on $C(G)$,
since $\sup_{g\in G}\sup_{||f||_\infty=1} \big|\iint d
\overline{\mu(a)}d\mu(b)f(agb^{-1})\big|=|\mu|^2(G)$.
It is also completely positive definite,
which can be easily proven using the GNS Theorem \ref{GNS}:
\begin{eqnarray}
(\I\otimes\Lambda_\mu)\varphi(g_1,g_2)&=&\iint d\overline{\mu(a)}d\mu(b)
\big\langle v_\varphi\big|\pi_\varphi(g_1,ag_2b^{-1}) v_\varphi\big\rangle\nonumber\\
&=& \iint d\overline{\mu(a)}d\mu(b)
\big\langle \pi_\varphi(e,a)^\dagger v_\varphi\big|\pi_\varphi(g_1,g_2)
\pi_\varphi(e,b)^\dagger v_\varphi\big\rangle\nonumber\\
&=& \big\langle \pi_\varphi(e,\mu)^\dagger v_\varphi\big|\pi_\varphi(g_1,g_2)
\pi_\varphi(e,\mu)^\dagger v_\varphi\big\rangle,
\end{eqnarray}
where $\pi_\varphi(e,\mu):=\int d\mu(g)\pi_\varphi(e,g)$.  Thus, $(\I\otimes\Lambda_\mu)\varphi$
is positive definite.

Map (\ref{Lmu}) can be further generalized:
\begin{equation}\label{LM}
\Lambda_{\mathfrak M} f(g):= \int\limits_{M(G)}d\mathfrak M(\mu)\,\big(\mu^*\ast f \ast\mu\big)(g),
\end{equation}
where $\mathfrak M$ is a positive measure on $M(G)$ with a finite total variation,
i.e. $\mathfrak M\in M\big(M(G)\big)$.
We conjecture that any CPD map from $C(G)$ to $C(G)$
is of this form
for some measure $\mathfrak M$, so that Eq. (\ref{LM}) is an analog
of the Kraus decomposition of a completely positive map.

\section{Conclusions}

The main conclusions are twofold. On one hand, this paper is directed to the audience of mathematicians
working in the area of harmonic analysis. We have formulated here the separability problem in purely abstract
terms of positive  definite functions on compact groups. To our knowledge this is a new problem in harmonic
analysis.  One may hope that some well established methods of harmonic analysis will help to get more insight
into the problem, solving it, at least partially. Several
generalizations call for immediate attention: applications to quantum groups being perhaps one of the most
fascinating ones. We hope that studies of entanglement within the harmonic analysis framework will open new
avenues here.

One the other hand, we expect that the harmonic analysis methods will help to study the physics of
separability and entanglement.  We expect to find new entanglement criteria, and get a better understanding of
the whole problem, by looking at concrete examples and applications of our theoretical results.  In particular,
finite groups (which have finitely many irreducible representations) of various types (nilpotent, solvable) seem
a rich source of interesting examples.

As seen above, passing from linear-algebraic entanglement criteria for
quantum states to the theory of positive-definite functions on compact
groups is somewhat inovolved and requires technical work.  We expect, on
the other hand, that, going in the opposite direction, the above general
results, when specialized to concrete groups (e.g. finite groups or
$SU(2)$) will directly yield physically interesting results.

\section*{Acknowledgements}
We would like to thank M. Bo\.zejko, P. Horodecki, M. Marciniak, P. So\l tan, and S. L. Woronowicz
for discussions.
We gratefully acknowledge the financial support of
EU IP  Programme ``SCALA'', ESF PESC Programme ``QUDEDIS'', Spanish MEC grants (FIS 2005-04627,
Conslider Ingenio 2010 ``QOIT''), and Trup Cualitat Generalitat de Catalunya.
JW was partially supported by the NSF grant DMS 9706915.

\end{document}